\documentclass[sigconf]{acmart}

\usepackage{balance}
\usepackage{bbm}
\usepackage{dblfloatfix} 
\usepackage{color}
\usepackage{xcolor}
\usepackage[disable]{todonotes}
\usepackage{graphicx}
\usepackage[lined, ruled, boxed, commentsnumbered,noend]{algorithm2e}
\usepackage{multirow}
\usepackage{amsmath,amssymb}
\usepackage{booktabs}
\usepackage{verbatim}
\usepackage{subcaption}

\newtheorem{informal}{Informal Problem}

\copyrightyear{2019}
\acmYear{2019} 
\setcopyright{iw3c2w3}
\acmConference[WWW '19]{Proceedings of the 2019 World Wide Web Conference}{May 13--17, 2019}{San Francisco, CA, USA}
\acmBooktitle{Proceedings of the 2019 World Wide Web Conference (WWW '19), May 13--17, 2019, San Francisco, CA, USA}
\acmPrice{}
\acmDOI{10.1145/3308558.3313449}
\acmISBN{978-1-4503-6674-8/19/05}

\begin{document}
	
\author{Subhabrata Mukherjee}
\authornote{\vspace{-.5em}Work done at Max Planck Institute, Germany prior to joining Amazon.}
\affiliation{%
	\institution{Amazon, USA}
}
\email{subhomj@amazon.com}

\author{Stephan G\"{u}nnemann}
\affiliation{%
	\institution{Technical University of Munich, Germany}
}
\email{guennemann@in.tum.de}

\title{GhostLink: Latent Network Inference for\\ Influence-aware Recommendation}

\begin{abstract}
Social influence plays a vital role in shaping a user's behavior in online communities dealing with items of fine taste like movies, food, and beer.
For online recommendation, this implies that users' preferences and ratings are influenced due to other individuals. 
Given only time-stamped reviews of users, can we find out who-influences-whom, and characteristics of the underlying influence network? Can we use this network to improve recommendation? 
 
 While prior works in social-aware recommendation have leveraged social interaction by considering the {\em observed} social network of users, many communities like Amazon, Beeradvocate, and Ratebeer \textit{do not have explicit user-user links}. 
Therefore, we propose GhostLink, an unsupervised probabilistic graphical model, to automatically \textit{learn} the latent influence network underlying a review community -- given only the temporal traces (timestamps) of users' posts and their content.  
Based on extensive experiments with four real-world datasets with $13$ million reviews, we show that GhostLink improves item recommendation by around $23\%$ over state-of-the-art methods that do not consider this influence. As additional use-cases, we show that GhostLink can be used to differentiate between users' latent preferences and influenced ones, as well as to detect influential users based on the learned influence~graph.

\end{abstract}

%
%



%
%

	\vspace{-30em}

\maketitle

\section{Introduction}

Traditional works in recommender systems that build upon collaborative filtering \cite{koren2011advances} exploit that similar users have similar rating behavior and facet preferences. 
Recent works use review content~\cite{mcauleyrecsys2013, wang2011, mukherjeeSDM2014} and temporal patterns~\cite{Gunnemann2014,DBLP:conf/kdd/MukherjeeGW16,DBLP:conf/www/GunnemannGF14} to extract further cues. All of these works assume the users to behave independently of each other. 
In a social community, however, users are often influenced by the activities of their friends and peers. How can we detect this influence in online  communities?

 \begin{figure}[!htbp]
	\centering
	\includegraphics[scale=0.45]{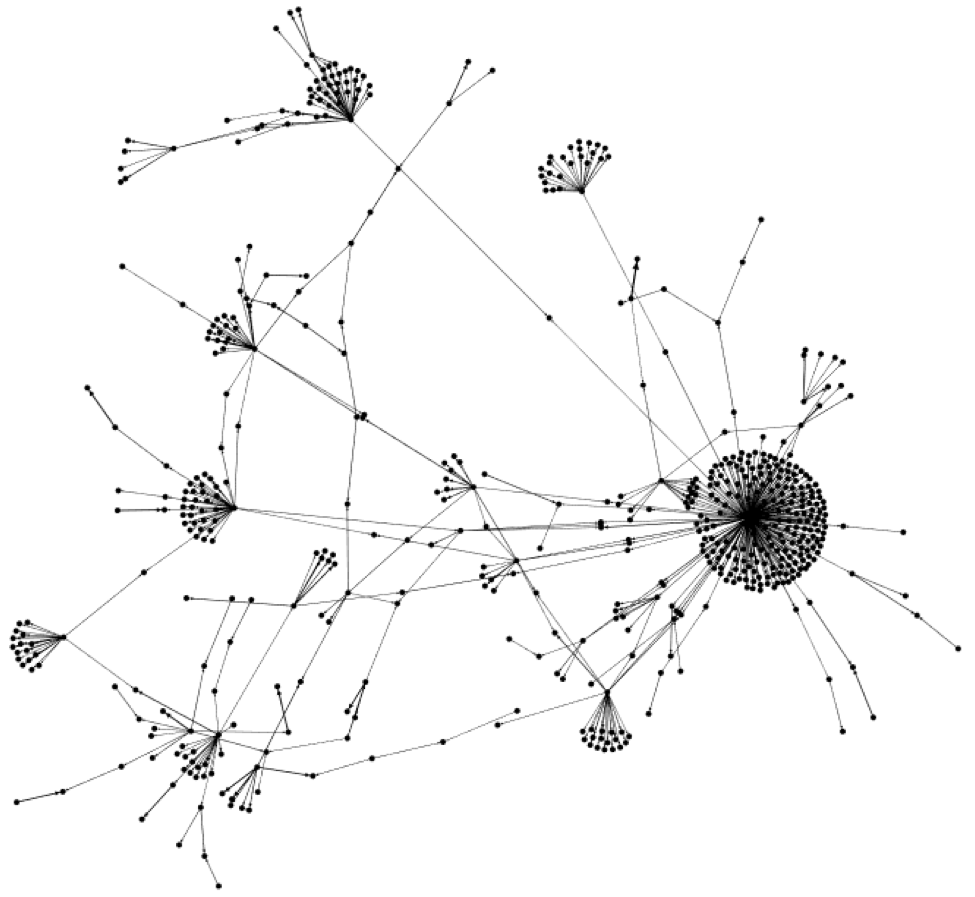}
	\caption{\small Given only timestamped reviews of users in the Beeradvocate community without any explicit user-user link/interaction, GhostLink extracts this latent influence network (of top $K$ influencers) based on opinion conformity. This is compactly represented by a Maximum Weighted Spanning Forest (MWSF) preserving $99.4\%$ of the influence mass from $73.4\%$ of the edges of the inferred influence network depicting a tree-like structure of influence. }
	\label{fig:infl-graph}
	\vspace{-1em}
\end{figure}

One way to answer this question is to exploit the \textit{observed} social network or interaction of users --- like friend circles in Facebook, the follow graph in Twitter, and trust relations in Epinion. Recent  works~\cite{Tang:2012:MDM:2124295.2124309,Tang:2013:ELG:2540128.2540519,DBLP:journals/datamine/LiuTHY12,DBLP:conf/sdm/ZhangYWSZ17, DBLP:journals/ida/MeiYSM17, DBLP:journals/jidm/FelicioPAAP16, Ye:2012:ESI:2348283.2348373, 7944514, Krishnan:2010, DBLP:conf/ecai/HuangCGSY10} 
 \text{leverage} such \textit{explicit} user-user relations or the observed social circle to 
propose \text{social-network} based recommendation. 
Similarly, in the field of citation \text{networks}, \cite{DBLP:conf/icml/DietzBS07}  attempt to extract citation influence given the {\em explicit network} of who-cited-whom. 


However, there is one big catch: many online review communities like Amazon or Beeradvocate do {\em not} have any explicit social network -- thus, making the above methods not applicable. Can we infer the influence network based on other signals in the data? 

While some recent works~\cite{Guo:2014:RTE:2554850.2554878, Lin:2014:PNR:2535053.2535249, Ma:2013:ESI:2484028.2484059, Ma:2011:RSS:1935826.1935877} model implicit relationships, they are limited to the historical rating behavior of users, ignoring the textual information. 
Similarly, works in information diffusion over latent networks model temporal traces ignoring the textual information~\cite{NetInf, Connie, NetRate, NetInfluence} and make some strong assumptions like a homogeneous network with static transmission rates. These techniques being agnostic of the context fail to capture complex interactions resulting in sparse networks. Some recent works on text-based diffusion~\cite{Wang2014, Du2013, HawkesTopic} model context. However, they also make some strong assumptions regarding the topics of diffusion being known apriori and the network being explicit. Most importantly, none of these works are geared for item recommendation, nor do they study the characteristics of review communities.

In contrast, in this work, we leverage opinion conformity based on writing style as an indication of influence: where a user echoes/ copies facet descriptions from peers (called influencers) across multiple items. This is a common setting for communities dealing with items of fine taste like movies, beer, food and fine arts where users often co-review multiple items.
Our informal goals are:
\vspace{-0.5em}
\begin{informal}
Given only timestamped reviews of users in online communities, \textbf{extract} the underlying influence network of who-influences-whom based on opinion conformity, and \textbf{analyze} the characteristics of this influence network.
\vspace{-0.5em}
\end{informal}

\begin{informal}
\textbf{Leverage} the implicit social influence (network) to improve item rating prediction based on peer activities.
\vspace{-0.5em}
\end{informal}

%
%

%

{\small
	\begin{table}
		\begin{tabular}{p{8cm}}
			\toprule

			{\bf Amazon Movies}\\
			\hspace{0.4em}{\bf U1: }style intense pretentious non-linear narrative rapid editing\\
			\hspace{0.4em}{\bf U2: }non-linear narrative crazy flashback scene randomly interspersed\\
			\midrule
			
			{\bf Beeradvocate}\\
			\hspace{0.4em}{\bf U1: } cloudy reddish amber color huge frothy head aroma spicy\\
			\hspace{0.4em}{\bf U2: } hazy golden amber dissipating head earthy aroma pepper clove\\
			
			\bottomrule
		\end{tabular}
		\caption{\small Sample (influenced) review snippets extracted by Ghost- Link from two communities: user U1's review is influenced by U2.}
		\label{tab:snapshot}
		\vspace{-3em}
	\end{table}
}

To answer these questions, we propose GhostLink, an unsupervised probabilistic graphical model, that automatically extracts the (latent) influence graph 
underlying a review community. 

{\em Key idea and approach:} Consider two users reviewing a movie in a review community. The first user expressed fascination for the movie's `non-linear narrative style', `structural complexity', and `cinematography' as outlined in the content of her review. Later, following this review, a second user also echoed similar concepts such as `seamless narrative', `style', and `matured cinematography'. That is, the second review closely resembles the first one {\em conceptually} in terms of \emph{facet descriptions} -- not simply by using the same words. While for a \emph{single} item this could be simply due to chance, {\em a repeated occurrence of this pattern across multiple items} -- where the second user reviewed an item some time {\em after} the first user echoing \emph{similar facet descriptions} -- gives an indication of influence. A user could be influenced by several users for different facets in her review. GhostLink models this notion of multiple influence common in communities dealing with items of fine taste like movies, food and beer where users often co-review multiple items.  Table~\ref{tab:snapshot} shows a snapshot of (influenced) review snippets extracted by GhostLink.

Based on this idea, we propose a probabilistic model that exploits the facet descriptions and preferences of users --- based on principles similar to Latent Dirichlet Allocation --- to learn an influence graph. 
Since the influencers for a given user and facets are unobserved, all these aspects are learned solely based on their review content and their temporal footprints (timestamps).


Figure~\ref{fig:infl-graph} \todo{please remove the colors! pick only the top-k nodes; remove the facet specfic graph!!!}shows such an influence graph  extracted by GhostLink from the Beeradvocate data. Analyzing these graphs gives interesting insights: There are only a few users who influence most of the others, 
and the distribution of influencers vs. influencees follows a power-law like distribution. Furthermore, most of the mass of this influence graph is concentrated in giant tree-like component(s). 
We use such influence graphs to perform influence-aware item recommendation; and we show that GhostLink outperforms state-of-the-art baselines that do not consider latent influence. Moreover, we use the influence graph to find influential users and to distinguish between users' latent facet preferences from that of induced/influenced ones.
%
%
%
Overall, our contributions are:
{\setlength{\leftmargini}{12pt}
\vspace*{-1mm}
\begin{itemize}
	\item {\bf Model:} We propose an unsupervised probabilistic generative model GhostLink based on Latent Dirichlet Allocation to learn a latent influence graph in online communities without requiring explicit user-user links or a social network. This is the first work that solely relies on timestamped review data. 
	\item {\bf Algorithm:} We propose an efficient algorithm based on Gibbs sampling~\cite{Griffiths02gibbssampling} to estimate the hidden parameters in GhostLink that empirically demonstrates fast convergence.
	\item {\bf Experiments:} We perform large-scale experiments in four communities with $13$ million reviews, $0.5$ mil.\ items, and $1$ mil.\ users where we show improved recommendation for item rating prediction by around $23\%$ over state-of-the-art methods. Moreover, we analyze the properties of the influence graph and use it for use-cases like finding influential members in the community.
\end{itemize}}


\section{GhostLink: Influence-Facet Model}

Our goal is to learn an influence graph between users based on their review content (specifically, overlap of their facet preferences) and timestamps only. The underlying assumption is that when a user $u$ is influenced by a user $v$, $u$'s facet preferences are influenced by the ones of $v$. Since the only signal  is the textual information of the reviews -- and their inferred latent facet distributions (also known as topic distributions in the context of LDA) -- we argue that influence is reflected by the used/echoed words and facets. 

While classical user topic models assume that each word of a document is associated with a topic/facet that follows the user's preference, we assume that the topic/facet of each word might be based on the preferences of \emph{other} users as well -- the influencers.
Inspired by this idea, we first describe the generative process of GhostLink followed by the explanation of the inference procedure.

\noindent\textbf{Generative Process.}
Consider a corpus $D$ of reviews written by a set of users $U$ at timestamps $T$ on a set of items $I$. The subset of reviews for item $i\in I$ is denoted with $D_i\subseteq D$. Let $d \in D_i$ be a review on item $i \in I$,  we denote with $u_{d}$ the user and with $t_d$ the timestamp of the review. 
All the reviews on an item $i$ are assumed to be ordered by timestamps.
Each review $d$ consists of a sequence of $N_d$ words denoted by $d=\{w_1,\ldots ,w_{N_d}\}$, where each word is drawn from a vocabulary $W$ having unique words indexed by $\{1, \dots, W\}$. The number of latent facets/topics corresponds to~$K$.

In most review communities, a user browses through other reviews on an item before making a decision (say) at time $t$.  Therefore, the set of users and corresponding reviews that could potentially influence the given user's perspective on an item $i$ consists of all the reviews $d'$ written at time $t_{d'} < t$. We call the corresponding set of users -- the potential influence set $ IS_{u,i}=\{u'\in U\mid \exists d,d'\in D_i:u=u_d \wedge u'=u_{d'} \wedge   t_{d'} < t_d\} $ for user $u$ and item~$i$.

In our model, each user is equipped with a latent facet preference distribution $\theta_u$, whose elements $\theta_{u,k}$ denote the preference of user $u$ for facet $k \in K$. That is, $\theta_u$ is a $K$-dimensional categorical distribution; we draw it according to $\theta_u \sim Dirichlet_K(\alpha)$
with concentration parameter $\alpha$. These distributions later govern the generation of the review text (similar to LDA).

Furthermore, for each user an influence distribution $\psi_u$ is considered, whose elements $\psi_{u,v}$ depict the influence of user $v$ on user $u$. That is $\psi_u$ represents a $U$-dimensional categorical distribution -- \emph{and all $\psi_*$ together build the influence graph we aim to learn} (see also Sec.~\ref{sec:network}). Similar to above we define $\psi_u \sim Dirichlet_U(\rho)$.

When writing a review, a user $u$ can decide to write an original review based on her latent preferences $\theta_u$ --- or be influenced by someone's perspective from her influence set for the given item; that is, using the preferences of $\theta_v$ for some $v\in IS_{u,i}$.
Since a user might not be completely influenced by other users, we allow each word of the review to be either original or based on other influencers. 
More precise: For each word of the review, we consider a random variable $s\sim Bernoulli(\pi_u)$ that denotes whether it is original or based on influence, where $\pi_u$ intuitively denotes the `vulnerability' of the user $u$ to get influenced by others. 

If $s=0$ the user uses her own latent facet preferences. That is, following the idea of standard topic models, the latent facet for this word is drawn according to $z\sim  Categorical(\theta_u)$.
If $s=1$, user $u$ writes under influence. In this case, the user chooses potential influencer(s) $v$ according to the strength of influence given by $\psi_u$. Since for a specific item $i$, the user $u$ can only be influenced by users in $ IS_{u,i}$ who have written a review {\em before} her, we write
$v  \sim Categorical(\psi_u\cap  IS_{u,i})$
to denote restriction of the domain of $\psi_u$ to the currently considered influence set.

\begin{figure}[htbp]
	\centering
	\includegraphics[scale=0.6]{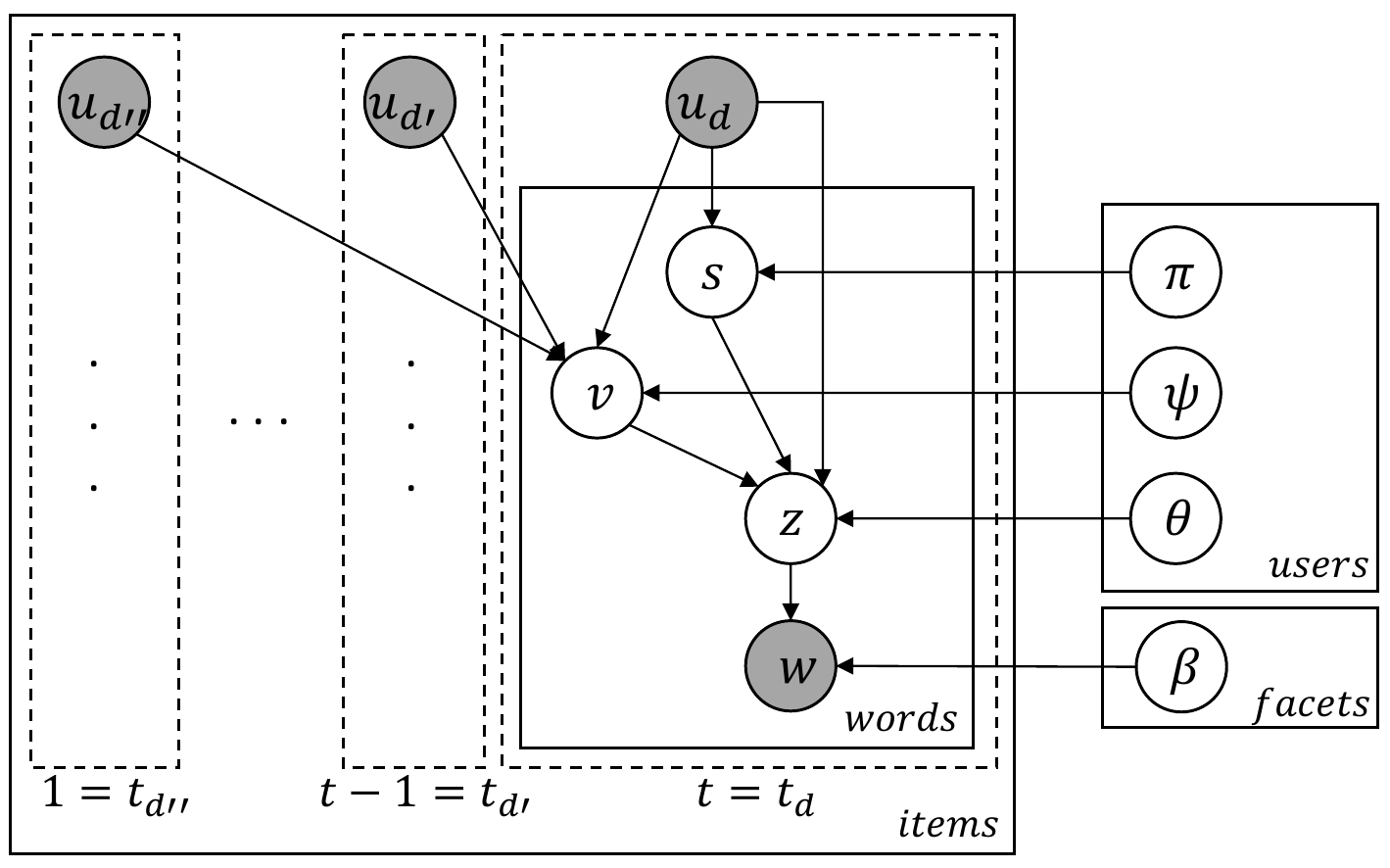}
	\vspace{-2em}
	\caption{Plate diagram for the generative process. Each dashed box indicates a single review.}
	\vspace{-1em}
	\label{fig:model}
\end{figure}

Given the sampled user $v$, the latent facet for this word should be drawn according to $v$'s preferences. Now, we are faced with a modeling choice. 
We can use the influencer's overall facet distribution $\theta_v$ (as is used in citation-network based topic models). However, by using $\theta_v$, one considers $v$'s \emph{generic} facet distribution -- which might be very unrelated to the item under consideration. That is, while $v$ might prefer specific facets, in his actual review $d'$ about item $i$ these facets might have not been used. And accordingly, since the user $u$ only sees the observed review $d'$ -- and not the latent facet distribution of $v$ --
the user $u$ cannot be influenced by facets which have not been considered.
Thus, in our model, instead of considering the influencer's (gene\-ric) facet distribution $\theta_v$, we consider the facet distribution that the influencer has actually {\em used} for writing his review $d'$ for the given item $i$. Since the review of the user $v$ has already been generated (otherwise the user would not be in $IS_{u,i}$), the used facets $z$ for each word of his review are known. Thus, instead of considering $\theta_v$, we consider the `observed' facet distribution based on the actual review, denoted with $\tilde{\theta}_{v}^{d'}$. Given this distribution, we sample $z\sim  Categorical(\tilde{\theta}_{v}^{d'})$. Since the model samples an influencer for each facet, a user can have {\em multiple influencers} corresponding to multiple facets in his review.

In summary, in the above process, the user $u$ either draws the facet $z$ from $\theta_u$ (if $s=0$) or $\tilde{\theta}_{v}^{d'}$ (if $s=1$).
Given facet $z$, we draw the actual word $w\sim Categorical(\beta_z)$ following the generative process of Latent Dirichlet Allocation \cite{Blei2003LDA}. As usual, $\beta_z\sim Dirichlet_W(\gamma)$ denotes corresponding per-facet word distributions.

Overall, the user's review can be regarded as being generated by a \emph{mixture} of her latent preferences and preferences of her influencers. Algorithm \ref{algo:1} summarizes the generative process, and the graphical model is illustrated in Figure \ref{fig:model}, where we indicated with $u_d$ the (observed) user for each review. 

\begin{algorithm}[t]
	\SetAlgoLined
	\DontPrintSemicolon
	{\small
		1. Draw $\theta_u \sim \text{Dirichlet}_K(\alpha)$ // latent facet preference of each user \;
		2. Draw $\psi_u \sim \text{Dirichlet}_U(\rho)$ // influencer distribution of each user \;
		3. Draw $\pi_u \sim \text{Beta}(\eta)$ // vulnerability of each user to be influenced  \;
		4. Draw $\beta_k \sim \text{Dirichlet}_W(\gamma)$ // word distribution of each facet  \;
		\For {each item $i\in I$} {
			\For {each review $d \in D_i$ on $i$ at time $t$ by user $u$} {
				\For {each word $w$ in $d$} {
					5. Draw $s \sim \text{Bernoulli}(\pi_u)$ \;
					\If {$s=0$} {
						6. $\theta' = \theta_u$ // use latent facet preference of user \;
					}
					\If {$s=1$} {
						7. Draw $v  \sim \text{Categorical}(\psi_u\cap IS_{u,i})$ \;
						8. $\theta' = \tilde{\theta}_{v}^{d'}$ // use the influencer's facet preference \;
						/* where $\tilde{\theta}_{v}^{d'}$  is the facet distribution {\em used} by $v$ for review $d'$ written at time $t' < t$ */ \;
					}
					9. Draw $z \sim \text{Categorical}(\theta')$ \;
					10. Draw $w \sim \text{Categorical}(\beta_z)$ \;
				}
			}
		}
	}	
	\caption{Generative process for influence -- facet model.}
	\label{algo:1}
\end{algorithm}



\section{Joint Probabilistic Inference} 

We now describe the inference procedure for GhostLink. That is, given the set of all reviews (and their timestamps), we aim to infer the latent variables.
To not clutter notation, we drop the indices of variables when it is clear from context (e.g.\ $\theta_u$ is abbreviated as $\theta$).

Let $S, V, Z$ be the set of all latent variables corresponding to the influence variables $s$, influencers $v$, and facets $z$. Let $W'$ denote the set of latent variables corresponding to the observed words, and $U'$ the set of latent variables corresponding to the observed users\footnote{Note that $U'$ refers to the latent variable attached to each review that `stores' the user information. Thus, a user might appear multiple times in $U'$ since she might have written reviews on multiple items. Similar for $W'$.}.
The joint probability distribution of our model is:
{\small
\vspace{-0.5em}
\begin{multline}
P(S, V, Z, W', \theta, \beta, \psi, \pi | U'; \alpha, \gamma, \rho, \eta)  \propto\\
 \prod_{u \in U} \big ( P(\pi; \eta)\cdot P(\psi; \rho) \cdot P(\theta; \alpha)  \big ) \cdot 
 \prod_{k \in K} P(\beta_k; \gamma)\cdot \\
 \prod_{i \in I} \prod_{d \in D_i} \bigg ( \prod_{s \in S}  
    P(s | \pi_{u_d}) \cdot 
  \prod_{v \in V} P(v | \psi_{u_d})^{\mathbb{I}(s=1)}
   \cdot \\
   \prod_{z \in Z} \big(   P (z | \theta_{u_d})^{\mathbb{I}(s=0)}  \cdot  P(z | \tilde{\theta}_v^{d'})^{\mathbb{I}(s=1)} \big )
  \cdot    \prod_{w \in W'}  P(w | \beta_z) \bigg)
\end{multline}
}

Since exact inference is intract\-able, we have to resort to approximate inference. 
For this purpose, we perform Collapsed Gibbs Sampling \cite{Griffiths02gibbssampling}. In Gibbs sampling, the conditional distribution for each hidden variable is computed based on the current assignment of the other hidden variables. The values for the latent variables are sampled repeatedly from this conditional distribution until convergence. In our problem setting we have three sets of latent variables corresponding to $S, V$ and $Z$ respectively -- the remaining variables $\theta, \beta, \psi, \pi$ are marginalized out (collapsed).

Given the current assignment of random variables, we use the  shortcuts:
$n(u,s)$ denotes the count of words written by $u$ with influence variable $s \in \{0,1\}$.  $n(u, v, s=1)$ denotes the count of words written by $u$ under influence from $v$ (i.e. $s=1$) in the community across all items and facets.  $n(u, z, s=0)$ denotes the number of times $u$ wrote facet $z$ for any word based on her latent preferences (i.e. $s=0$). 
 $n(v_{d'}, z)$ denotes the count of facet $z$ in review $v_{d'}$, and $n(z,w)$ denotes the number of times word $w$ is used with facet $z$. 

\noindent\textbf{Collapsing.} We first marginalize out the remaining variables as mentioned above. Exploiting conjugacy of the Categorical and Dirichlet distributions, we can integrate out $\pi$, $\psi$, $\theta$, and $\beta$ from the above distribution to obtain the four  posterior distributions\\[-4mm]

{\small
\[
P(S| U'; \eta )=
\frac{\Gamma(\sum_{s} \eta) \prod_{s} \Gamma(n(u,s) + \eta)}{\prod_{s} \Gamma(\eta) \sum_{s} \Gamma(n(u,s) + 2 \cdot \eta) }
\]
\[
P(V | U', S; \rho)= \frac{\Gamma(\sum_{v} \rho) \prod_{v} \Gamma(n(u,v,s=1) + \rho)}{\prod_{v} \Gamma(\rho) \sum_{v} \Gamma(n(u, v, s=1) + U \cdot \rho) }
\]
\[
P(Z| U', S, V; \alpha)=\frac{\Gamma(\sum_{z} \alpha) \prod_{z} \Gamma(n(u,z, s=0) + \alpha)}{\prod_{z} \Gamma(\alpha) \sum_{z} \Gamma(n(u,z, s=0) + K \cdot \alpha) }
\]
\[
P(W' | Z; \gamma)=\frac{\Gamma(\sum_{w} \gamma) \prod_{w} \Gamma(n(z, w) + \gamma)}{\prod_{w} \Gamma(\gamma) \sum_{w} \Gamma(n(z, w) + W \cdot \gamma) }
\]
} where $\Gamma$ denotes the Gamma function\footnote{The derivation of the following equations --- for integrating out latent variables from the joint distribution exploiting Multinomial-Dirichlet conjugacy and Gibbs Sampling updates  --- follow from the standard principles of Latent Dirichlet Allocation, and, therefore, details have been omitted for space.}.

\noindent
\textbf{Gibbs sampling.}
Given the above, the joint probability distribution with conditional independence assumptions is:
\[
P(S, V, Z | U', W') \propto  P(S|U') \cdot P(V|S,U') \cdot P(Z|V,S,U') \cdot P(W' | Z)
\]

The factors on the right-hand side capture (in order): vulnerability of the user being influenced, potential influencers given the user, facet distribution to be used (latent or influenced), and subsequent words to be used according to the facet distribution chosen. We infer all the distributions using Gibbs sampling.

 Let the subscript $-j$ denote the value of a variable excluding the data at the $j^{th}$ position. The conditional distributions for Gibbs sampling for updating the latent variable $S$ --- that models whether the user is going to write on her own or under influence --- is:\\[-6mm]

{\small
\begin{align}
P(s_j=0| u_d, z,s_{-j}) \propto \frac{n(u_d, s_j=0) + \eta}{\sum_s n(u_d, s) + 2 \cdot \eta} \cdot \frac{n(u_d, z, s_j=0) + \alpha}{\sum_z n(u_d, z, s_j=0) + K \cdot \alpha} \label{eq2}\\
\tilde{P}(s_j=1| u_d, v_{d'}, z, s_{-j}) \propto \frac{n(u_d, s_j=1) + \eta}{\sum_s n(u_d, s) + 2 \cdot \eta} \cdot \frac{n(v_{d'}, z) + \alpha}{\sum_z n(v_{d'}, z) + K \cdot \alpha} \label{eq3}\\
P(s_j=1| u_d, z, s_{-j}) \propto {max}_{v_{d'} \in D_i : t_{d'} < t_d} \tilde{P}(s_j=1| u_d, v_{d'}, z, d,s_{-j}) \label{eq4}
\end{align}
\vspace{-1em}
}

\todo{I changed the above from argmax to max. Is this correct?}

The first factor in Equation \ref{eq2} and \ref{eq3} above models the probability of the user being influenced: as a fraction of how many facets the user wrote under influence ($s=1$), or otherwise ($s=0$), out of the total number of facets written. 
The second factor in Equation~\ref{eq2} models the user's propensity of using a particular facet based on her latent preferences (when $s=0$); whereas the second factor in Equation~\ref{eq3} models the probability of the user writing about a facet under influence (when $s=1$) from an earlier review on the given item. Note that in this case --- as the user is influenced by another user's review that appeared earlier in her timeline --- she adopted her influencer's {\em used facet distribution} to write about the given facet instead of her own latent facet preference distribution. 

Note that in the above question, we did not assume the influencers $v_d'$ to be given since this would lead to a very restrictive Gibbs sampling step. Instead, as shown in  Equation~\ref{eq4}, we sample the best possible influencer for a given facet to determine the probability for $s=1$. Accordingly, the influencer for a given user and word, when writing under influence, is updated as\\[-9mm]

{\small
\begin{multline}
\label{eq5}
	{v_{d'}}_j | u_d, s=1, z, {v_{d'}}_{-j} = {argmax}_{v_{d'} \in D_i : t_{d'} < t_d} \bigg(\\ \frac{n(u_d, v_{d'}, s=1) + \rho}{\sum_v n(u_d, v, s=1) + U \cdot \rho} \cdot \frac{n(v_{d'}, z) + \alpha}{\sum_z n(v_{d'}, z) + K \cdot \alpha} \bigg)
\end{multline}
}

The first factor above counts how many times $u$ has been influenced by $v$ on writing about any facet --- out of the total number of times $u$ has been influenced by any other member in the community. The second factor is the facet distribution {\em used} by $v$ in the review that influenced $u$'s current facet description.

Instead of computing Equations~\ref{eq4} and~\ref{eq5} separately, we perform the update of both --- $s_j$ and ${v_{d'}}_j | s_j=1$ ---  {\em jointly}, thereby, reducing the computation time significantly.

 The conditional distribution for sampling the latent facet $z$ is:\\[-6mm]
 
 {\small
 \begin{align}
 	P(z_j | u_d, s=0,z_{-j}) \propto \frac{n(u_d, z_j, s=0) + \alpha}{\sum_z n(u_d, z, s=0) + K \cdot \alpha} \cdot \frac{n(z_j, w) + \gamma}{\sum_w n(z_j, w) + W \cdot \gamma} \label{eq6}\\
 	P(z_j | u_d, v_{d'}, s=1, z_{-j}) \propto \frac{n(v_{d'}, z_j) + \alpha}{\sum_z n(v_{d'}, z) + K \cdot \alpha} \cdot \frac{n(z_j, w) + \gamma}{\sum_w n(z_j, w) + W \cdot \gamma} \label{eq7}
 \end{align}
 \vspace{-1em}
}

The first factor in the above equations models the probability of using the facet $z$ under the user's (own) latent preference distribution  (Equation~\ref{eq6}), or adopting the influencer's used facet distribution (Equation~\ref{eq7}). The second factor counts the number of times facet $z$ is used with word $w$ --- out of the total number of times it is used with any other word.

\subsection{Overall Processing Scheme}
Exploiting the above results, the overall inference is an iterative process consisting of the following steps. We sort all reviews on an item by timestamps. For {\em each word} in each review on an item:
\begin{enumerate}
	\item Estimate whether the word has been written under influence, i.e. compute $s$ using Equations~\ref{eq2} - \ref{eq4} keeping all facet assignments fixed from earlier iterations. 
	\item In case of influence (i.e. $s=1$), an influencer $v$ is jointly sampled from the previous step.
	\item Sample a facet for the word using Equations~\ref{eq6} and ~\ref{eq7} keeping all influencers and influence variables fixed.
\end{enumerate}
The process is repeated until convergence of the Gibbs sampling process (i.e. the log-likelihood of the data stabilizes).

\subsection{Example}

Consider a set of reviews written by three users in the following time order: first Adam, then Bob, then Sam (see Table \ref{tab:example}). The table also shows the {\em current} assignment of the latent variables $z$ and $s$. The goal is to {\em re-sample} the influence variables. For ease of explanation, we ignore the concentration parameters of the Dirichlet distribution in the example and we ignore the subscript $-j$ from the variables. That is, we do not exclude the current state of the own random variable as in Gibbs sampling. 

Similar to before, let $n(u, s)$ be the number of tokens written by $u$ with influence variable as $s$, $n(d, z)$ be the total number of tokens with topic as $z$ in document $d$, $n(d)$ be the number of tokens in document $d$, and $n(u)$ be the total number of tokens written by $u$.

 For Adam we have $s = 0$ for each word. As he is the first reviewer, he has no influencers. For Bob, the influence variable $s$ w.r.t.\ the word `non-linear' is based on:\\[-8mm]

{\small
\begin{multline*}
\hspace*{-4mm}P(s_{\text{`non-lin'}}\! =\! 0 | u\!=\!Bob, z\!=\!z_2)  
\propto \frac{n(u\!=\!Bob, s\!=\!0)}{n(u\!=\!Bob)} \cdot \frac{n(u\!=\!Bob, z\!=\!z_2, s\!=\!0)}{n(u\!=\!Bob, s\!=\!0)}\\
= \frac{1}{2} \cdot \frac{0}{1} 
= 0\\
 P(s_{\text{`non-lin'}} = 1 | u=Bob, v = Adam, z=z_2, v_d=d_1)  \\
 \propto \frac{n(u=Bob, s=1)}{n(u=Bob)} \cdot \frac{n(z=z_2, v_d=d_1)}{n(v_d=d_1)}
 =\frac{1}{2} \cdot \frac{2}{3} 
 = \frac{1}{3}
\end{multline*}
}

Therefore, Bob is more likely to write `non-linear' being influenced by Adam's review than on his own. 
Similarly, for Sam:

{\small
	\begin{align*}
	P(s_{\text{`non-lin'}}=0 | u=Sam, z=z_2) \propto \frac{1}{2} \cdot 1 = \frac{1}{2}
	\end{align*}
}

Note the higher probability compared to the one of Bob since Sam uses further terms (i.e. `thriller') which also belong to facet $z_2$ that he wrote uninfluenced. For the case $s=1$, we would obtain:

{\small
\begin{align*}
	P(s_{\text{`non-lin'}}=1 | u=Sam, v=Adam, z=z_2, v_d=d_1) &\propto \frac{1}{2} \cdot \frac{2}{3} =  \frac{1}{3}\\
	P(s_{\text{`non-lin'}}=1 | u=Sam, v=Bob, z=z_2, v_d=d_2) &\propto \frac{1}{2} \cdot \frac{1}{2} = \frac{1}{4}	
\end{align*}
}

As seen, Sam is more likely influenced by Adam's review, rather than by Bob's, when considering facet $z_2$, since  $d_1$ has a higher concentration of $z_2$. It is worth noting that the probability of the influence variable $s$ depends only on the facet, and not the exact words. Our model captures semantic or facet influence rather than just capturing lexical match.
Overall, however, Sam is likely to write `non-linear' on his own rather than being influenced by someone else since $P(s_{\text{`non-lin'}}=0|...)$ is larger.

While the above example considers a single item, in a community setting --- especially for communities dealing with items of fine taste like movies, food and beer where users co-review multiple items --- such statistics are aggregated over several other items. This provides a stronger signal for influence when {\em a user copies/echoes similar facet descriptions from a particular user across several items}. 
Our algorithm, therefore, relies on three main factors to model influence and influencer in the community:
{\setlength{\leftmargini}{12pt}
\begin{itemize}
	\item [a)] The vulnerability of a user $u$ in getting influenced, modeled by $\pi$ and captured in the counts of $n(u,s)$.
	\item [b)] The textual focus of the influencing review $v_d$ by $v$ on the specific facet ($z$), modeled by $\theta$ and captured in the counts of $n(v_d,z)$; as well as how many times the influencer $v$ influenced $u$, modeled by $\psi$ and captured in counts of $n(u,v,s=1)$ --- aggregated over all facets and items they co-reviewed.
	\item  [c)] The latent preference of $u$ for $z$, modeled by $\theta_u$ and captured in the counts of $n(u,z,s=0)$.
\end{itemize}}


{\small
	\begin{table}[!t]
		\begin{tabular}{llllll}
			\toprule
			\textbf{Reviewer} & \textbf{Document} & \textbf{time} & \textbf{Word} &\textbf {Facet} & \textbf{Influence (s=\ )}  \\
			Adam & $d_1$ & 0 & action & $z_1$ & 0\\
			&&			& non-linear & $z_2$ & 0\\
			&&			& narrative & $z_2$ & 0\\\midrule
			Bob & $d_2$ & 1 & action & $z_1$ & 0\\
			&			& 	&  non-linear & $z_2$ & 1\\\midrule
			Sam & $d_3$ & 2 & non-linear & $z_2$ & 1\\
			&			&	 & thriller & $z_2$ & 0\\ 
			\bottomrule
		\end{tabular}
		\caption{Example to illustrate our method.}
		\label{tab:example}
		\vspace{-2.5em}
	\end{table}
}

\subsection{Fast Implementation}
\label{subsec:fast}
In the above generative process, we sample a facet for {\em each token/word} in a given review. Thus, we may sample different facets for the same word present multiple times in a review. While this makes sense for long documents where a word can belong to multiple topics, for short reviews it is unlikely that the same word is used to represent different facets. Therefore, we reduce the time complexity by sampling a facet for {\em each unique token} present in a review. We modify our sampling equations to reflect this change.

In the original sampling equations, we let {\em each token} contribute $1$ unit to the counts of the distribution for estimation. Now, {\em each unique token} contributes $c$ units corresponding to $c$ copies of the token in the review. As we sample a value for a random variable during Gibbs sampling, ignoring its current state, we also need to discount $c$ units for the token (instead of $1$) to preserve the overall counts. All the sampling equations are modified accordingly.


\subsection{Constructing the Influence Network}\label{sec:network}
Our inference procedure computes values for the latent variables $S$, $V$, $Z$, and corresponding distributions. Using these, our objective is to construct the influence network given by $\psi$:


{\small
	\begin{equation*}
	\psi_{u, v} = \frac{n(u, v, s=1) + \rho}{\sum_v n(u, v, s=1) + U \cdot \rho}
	\end{equation*}
}

The above counts the number of facet descriptions $n(u, v, s=1)$  that are copied by $u$ from $v$  (with $s=1$ depicting influence) out of the ones copied by $u$ from anyone else.  
Given $\psi$, we can construct a directed, weighted influence network $G = (U, E)$, where each user $u \in U$ is a node, and the edgeset $E$ is given by: 

{\small
	\vspace*{-1mm}
	\begin{equation}
	\label{eq:psi}
	E = \big\{ \{ v, u\} \ | \ \psi_{u,v} > 0, u \in U, v\in U  \big\}
	\end{equation}
	\vspace*{-1mm}
}

That is, there exists an edge from $v$ to $u$, if $v$ positively influences $u$ with the edge weight being $\psi_{u,v}$. 

Furthermore, GhostLink can distinguish between different facet preference distributions of each user. The \emph{observed} facet preference distribution $\theta^{obs}_u$ of $u$ is given by: 

{\small
	\begin{equation*}
	\theta^{obs}_{u,z} = \frac{n(u, z) + \alpha}{\sum_z n(u, z) + K \cdot \alpha}
	\end{equation*}
}

This counts the proportion of times $u$ wrote about facet $z$ out of the total number of times she wrote about any facet --- with or without influence. This distribution represents essentially the preferences as it is captured by a standard author-topic model~\cite{rosenzviUAI2004}, user-facet model~\cite{mcauleyrecsys2013}, and most of the other works using generative processes to model users/authors.

With GhostLink, however, we can derive even more informative distributions: 


{\small
	\begin{equation*}
	\theta^{latent}_{u,z} = \frac{n(u, z,s=0) + \alpha}{\sum_z n(u, z,s=0) + K \cdot \alpha}
	\end{equation*}
}
{\small
\begin{equation*}
\theta^{infl}_{u,z} = \frac{n(v=u,z,s=1) + \alpha}{\sum_z n(v=u,z,s=1) + K \cdot \alpha}
\end{equation*}
}


The distribution $\theta^{latent}_u$ intuitively represents a user's latent facet preference when \emph{not being influenced} from the community (i.e. $s=0$). In contrast, $\theta^{infl}_{u}$ captures the facet distribution of $u$ as an influencer, i.e. that she used to influence someone else.  That is, the latter one counts the proportion of times $u$ was chosen as the influencer (i.e. $v=u$, $s=1$) by another user in the community; or in other words when some other user copied 
from $u$.


\section{Item Rating Prediction using Influence Networks}
\label{sec:item-rating}

Our proposed method learns an influence network $\psi$ from the review data. We hypothesize that using this network helps to improve rating prediction. 
That is, our objective is to predict the rating $y'_{u, i,t}$ that user $u$ would assign to an item $i$ at time $t$ exploiting her latent social neighborhood given by $\psi_{u}$.  Since we know the actual ground ratings $y_{u,i,t}$, the performance for this task can be measured by the mean squared error: MSE = $\frac{1}{|{U, I}|} \sum_{u,i} (y_{u,i,t} - y'_{u,i,t})^2$.
\noindent Note that we use the rating data only for the task of rating prediction -- it has {\em not been used} to extract the influence graph. 

In the following, we describe the features we will create for  each review for the prediction task. We will analyze and compare their effects in our experimental study. Recap that each review $d$ consists of a sequence of words $\{w\}$ by $u$ on item $i$ at time $t$.

\noindent {\bf F1. Language model features} based on the review text:
Using the learned language model $\beta$, we construct $\langle F_w = log ( max_{z}\beta_{z,w}) \rangle$
of dimension $W$ (size of the vocabulary). That is, for each word $w$ in the review, we consider the value of $\beta$ corresponding to the best
facet $z$ that can be assigned to the word. We take the log-transformation of $\beta$ which empirically gives better results. 

\noindent {\bf F2. Rating bias features}: Similar to~\cite{mcauleyrecsys2013,koren2011advances}, we consider:
	(i) Global rating bias $\gamma_g$: Average rating $avg(\langle y \rangle)$ assigned by all users to all items.
(ii) User rating bias $\gamma_u$: Average rating $avg(\langle y_{u,.,.} \rangle )$ assigned by $u$ to all items.
	(iii) Item rating bias $\gamma_i$: Average rating $avg(\langle y_{.,i,.} \rangle )$ assigned by all users to item $i$.

\noindent {\bf F3. Temporal influence features}: Finally, we  exploit the temporal and influence information we have learned with our model.
	%
	
\noindent (i) Temporal rating bias $\gamma_r$: Average rating $avg(\langle y_{.,i,t} \rangle )$ assigned by all users to item $i$ {\em before} time $t$. This baseline considers the temporal trend in the rating pattern.
	%
		\noindent (ii)  Temporal influence from rating $\gamma_d$: Let  $\langle d_{t} \rangle $ be the set of reviews written by  users $\langle v_{d_{t}} \rangle$ before time $t$ on the  item $i$. Consider the influence of $v_{d'}$ on $u$, i.e.\ the variable $\psi_{u, v_{d'}}$, as learned by our model. 
	    The feature $avg(\langle \psi_{u, v_{d_t}} \cdot y_{v_{d_t},i,t}  \rangle )$ aggregates the rating of each previous user and her influence on the current user for item $i$ at time $t$ to model the influence of previous users' ratings on the current user's rating. This baseline combines the temporal trend and the social influence of earlier users' rating. 
		%
	 %
		\noindent (iii)  Temporal influence from context $\gamma_{dc}$: Consider the  review $d$ with the sequence of words $\langle w \rangle$. Let $s_w \in \{0,1\}$ be the influence variable sampled for a word $w$, and $v_w$ be the influencer sampled for the word when $s_w=1$, as inferred from our model. Also, let $y_{v_w}$ be the rating assigned by $v_w$ to the current item at time $t' < t$. Consider $\mathbb{I}(.)$ be an indicator function that is $1$ when its argument is true, and $0$ otherwise. We use:\\[-8mm]
		
		{\small
		\begin{equation*}
			\gamma_{dc} = \frac{1}{|d|} \sum_{w \in d}  \bigg( \mathbb{I}(s_w=1) \cdot \big(\psi_{u, v_{w}} \cdot y_{v_w}\big) +  \mathbb{I}(s_w=0) \cdot \gamma_u \bigg)
		\end{equation*}
		}
		  
		  For each word $w$ in the review $d$, if the word is written under influence ($s_w = 1$), we consider the influencer's rating and her influence on the current user given by $\psi_{u, v_{w}} \cdot y_{v_w}$. Otherwise ($s_w=0$), we consider the user's self rating bias $\gamma_u$. This is aggregated over all the words in the review. 
		  This baseline combines the temporal trend and context-specific social influence of earlier users' rating. 

Using different combinations of these features (see Sec.~\ref{sec:experiments}), we use Support Vector Regression~\cite{drucker97} from LibLinear with default parameters to
 predict the item ratings $y'_{u,i,t}$, using ten-fold cross-validation ({https://www.csie.ntu. edu.tw/~cjlin/liblinear/}).


{\small
	\begin{table}[t]
		\setlength{\tabcolsep}{1.4mm}
		\begin{tabular}{lrrrc}
			\toprule
			\bf{Dataset} & \bf{\#Users} & \bf{\#Items} & \bf{\#Reviews} & \bf{\#Years}\\
			\midrule
			\bf{Beer (BeerAdvocate)} & 33,387 & 66,051 & 1,586,259 & 16\\
			\bf{Beer (RateBeer)} & 40,213 & 110,419 & 2,924,127 & 13\\
			\bf{Movies (Amazon)} & 759,899 & 267,320 & 7,911,684 & 16\\
			\bf{Food (Amazon)} & 256,059 & 74,258 & 568,454 & 16\\
			\midrule
			\bf{TOTAL} & 1,089,558 & 518,048 & 12,990,524 & -\\
			\bottomrule
		\end{tabular}
		\caption{Dataset statistics.}
		\label{tab:statistics}
		\vspace{-3em}
	\end{table}
}

\section{Experiments}
\label{sec:experiments}
We empirically analyze various aspects of GhostLink, using four online communities in different domains: BeerAdvocate ({\tt \small beeradvo cate.com}) and RateBeer ({\tt \small ratebeer.com}) for {\em beer} reviews. Amazon ({\tt\small amazon.com}) for {\em movie} and {\em food} reviews.
Table~\ref{tab:statistics} gives an overview. 
All datasets are publicly available at {\tt \small http://snap.stanford.edu}. 
We have a total of $13$ million reviews from $1$ million users over $16$ years from all of the four communities combined. From each community, we extract the following quintuple for GhostLink $<$$userId, itemId,$ $timestamp,$ $rating, review$$>$. We set the number of latent facets $K=20$ for all datasets. The symmetric Dirichlet concentration parameters are set as: $\alpha=\frac{1}{K},\eta=\frac{1}{2}, \rho=\frac{1}{U}, \gamma=0.01$.\footnote{We did not fine-tune hyper-parameter $K$. It is possible to improve performance by considering the value of $K$ that gives the best model perplexity. Similarly, we consider symmetric Dirichlet priors for a simplistic model with less hyper-parameters to tune.}

Performance improvements of GhostLink over baseline methods are statistically significant at $99\%$ level of confidence determined by {\em paired sample t-test}.

\subsection{Likelihood, Smoothness, Fast Convergence}
\todo{as said in my email, TWO lines per plot!!!}
There are multiple sets of latent variables in GhostLink that need to be inferred during Gibbs sampling.
Therefore, it is imperative to show the resultant model is not only stable, but also improves log-likelihood of the data. A higher likelihood indicates a better model.  There are several measures to evaluate the quality of facet models; we use here the one from~\cite{wallach}:
$LL = \sum_d \sum_{j=1}^{N_d} log\ P(w_{d,j} | \beta; \alpha)$.

\begin{figure}[b!]
	\centering
	\includegraphics[width=0.5\linewidth]{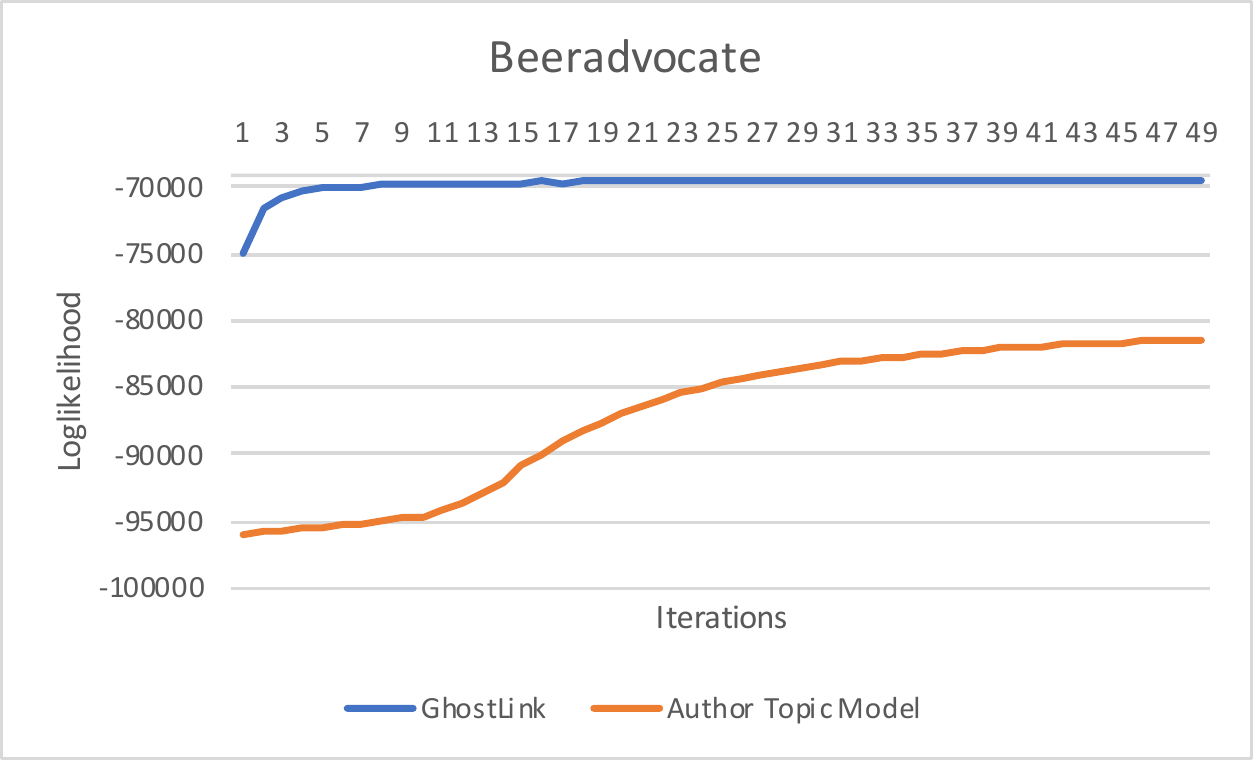}%
	\includegraphics[width=0.5\linewidth]{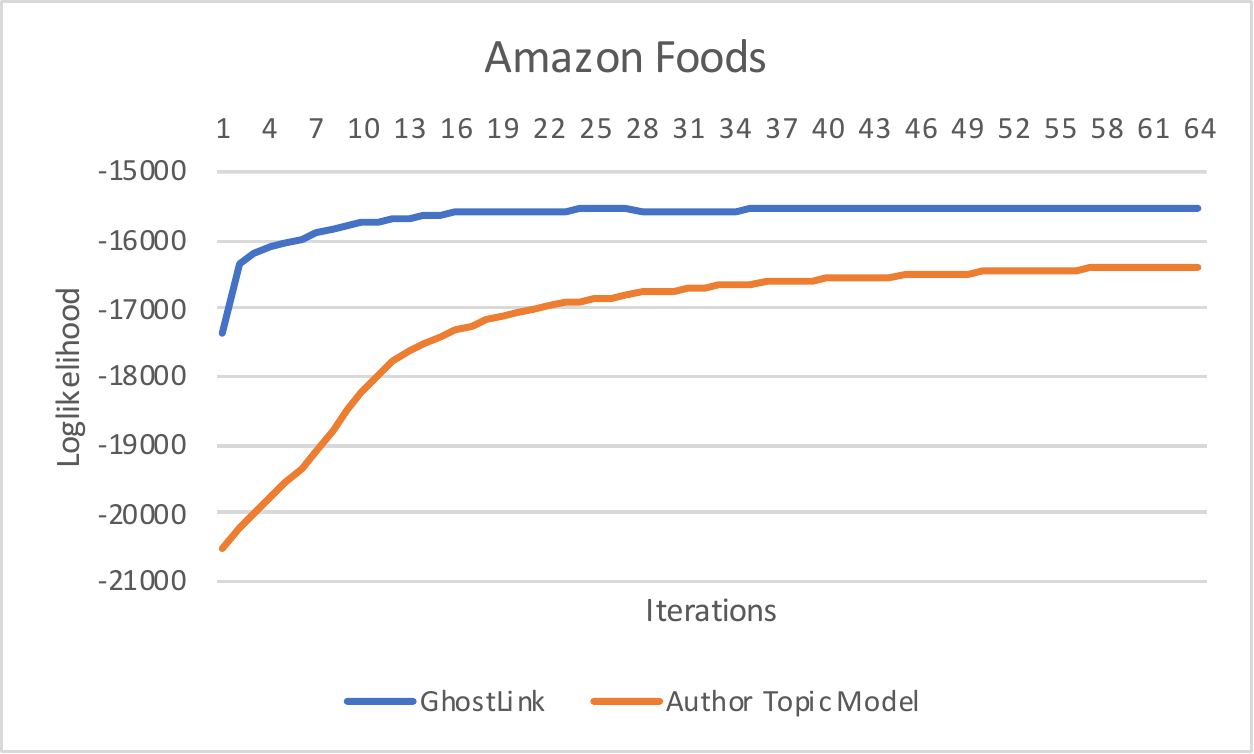}
	\caption{Log-likelihood of GhostLink and Author-Topic Model \cite{rosenzviUAI2004} per-iteration in Beeradvocate \& Amazon Foods.}
	\label{fig:log-likelihood}
\end{figure}

 {\small
	\begin{table}
		\centering
		\setlength{\tabcolsep}{1pt}
		\begin{tabular}{p{4cm}p{1cm}p{1cm}p{1cm}p{1cm}}
			\toprule
			& Beer &	Rate &	Amazon &	Amazon \\
			& advocate & beer &	 Foods &	 Movies\\\midrule
			GhostLink: Fast Implementation & 1.8 & 1.6 & 0.08 &	1.9\\
			GhostLink: Basic &	6 & 2.2	& 0.14 & 3.1\\
			\bottomrule
		\end{tabular}
		\caption{Run time comparison (in hours) till convergence between different versions of GhostLink.}
		\label{tab:time}
		\vspace{-3em}
	\end{table}
}

Figure~\ref{fig:log-likelihood} shows the log-likelihood of the data per iteration for the \todo[color=green]{for the datasets we have picked!!! all -- or mention here the names; and add a statment that similar results for other two darasets}Beeradvocate and Amazon Foods data. The plots for the other data\-sets are similar. 
We find that the learning is stable and has a {\em smooth} increase in the data log-likelihood {\em per iteration}. Empirically GhostLink also shows a fast convergence 
in around $10$ iterations. 

Table~\ref{tab:time} shows the run time comparison to convergence between the basic and fast implementation of GhostLink\footnote{Experiments are performed in: Intel(R) Xeon(R) CPU E5-2667 v3 @ 3.20GHz. Note that our Gibbs sampling based inference process is sequential and not distributed.}. The fast version uses two tricks: (i) instead of computing Equations~\ref{eq4} and~\ref{eq5} separately, it estimates --- $s_j$ and ${v_{d'}}_j | s_j=1$ ---  {\em jointly}, and (ii) it estimates facets for each {\em unique} token {\em once} as in Section~\ref{subsec:fast}.

We also compare the log-likelihood of our model to another generative model that is closest to our work namely, the Author-Topic Model~\cite{rosenzviUAI2004}. This work models documents (reviews) to have a distribution over authors, authors to have a distribution over topics, and topics to have a distribution over words. This model is easy to mimic in our setting by ignoring the notion of influence (i.e. setting $s=0$ as constant for all authors/users). Figure~\ref{fig:log-likelihood} shows the stark difference in log-likelihood of the two models where GhostLink considering influence ($s \in \{0,1\}$) performs much better than the baseline that ignores the effect of temporal influence ($ s \in \{0\}$).

\subsection{Influence-aware Item Rating Prediction}
Next, we show the effectiveness of GhostLink for item rating prediction. 
In Section~\ref{sec:item-rating}  we described the set of features and evaluation measure for this task.  Table~\ref{tab:MSE} compares the mean-squared error of GhostLink with all the baselines with {\em ten-fold cross validation} --- where we use $90\%$ of the data for training and $10\%$ for test with results averaged over $10$ such splits. 

We divide our baselines into four main categories. For each category, we chose the state-of-the-art system as a baseline that is the most representative of that category with all the features as applicable. Unavailability of explicit user-user links in our data renders many of the related works inapplicable to our setting. 

\noindent {\bf (A) Rating and Time-aware Latent Factor Models}: These baselines model users, items, ratings and their temporal dynamics but ignore the text or content of the reviews. For most of these baselines, we used the code repository from {http://cseweb.ucsd.edu/~jmcauley/code/}. Note that these models use the rating bias features (F2) from Section~\ref{sec:item-rating}. Since they do not model text or influence network, the other features are not applicable.

\noindent {\em (a) LFM}: This is the classical latent factor model based on collaborative filtering with temporal dynamics~\cite{KorenKDD2010} that considers ratings, latent facets, and time.

\noindent {\em (b) Community at uniform rate}: This set of models~\cite{McAuley2013,xiong2010temporal,Xiang2010} consider users and products in a community to evolve using a single global clock  with the different stages of community evolution appearing at uniform time intervals. So the preference for items evolves over time.

\noindent {\em (c) Community at learned rate}: This extends (b) by learning the rate at which the community evolves with time~\cite{McAuley2013}.

\noindent {\em (d) User at uniform rate}: This extends (b) to consider individual users and modeling users' progression based on their maturity and preferences evolving over time. The model assumes a uniform rate for evolution~\cite{McAuley2013}.

\noindent {\em (e) User at learned rate}: This extends (d) by allowing each user to evolve on their individual clock, so that the time to attain maturity varies for different users \cite{McAuley2013}.

\noindent {\bf (B) Text-aware Latent Factor Model}: Unlike the previous baselines, this model~\cite{mcauleyrecsys2013} considers text of the reviews along with the  latent factor models using collaborative filtering for item rating prediction. The authors learn topic/facet distributions from text using a generative model based on Latent Dirichlet Allocation, and tie them to the latent facet distributions learned from the collaborative filtering model based on users and ratings. All of these are jointly learned to minimize the mean squared error for item rating prediction. This is the strongest baseline for our work but ignores the notion of network influence. Note that this baseline uses the rating bias features (F2) and language model (F1) from Section~\ref{sec:item-rating}. The network influence features are not applicable.\footnote{We used their code publicly available at {http://cseweb.ucsd.edu/~jmcauley/code/}}. Also, note that the generative process in this work is similar to the Author Topic Model~\cite{rosenzviUAI2004} with the main difference of the former being tailored for item rating prediction.

\noindent {\bf (C) Network-aware Models}: We also experiment with two information diffusion based baselines~\cite{NetInfluence,NetInf}. Both models infer the latent influence network underlying a community based on only the temporal traces of activities (e.g., timestamps of users {\em posting} reviews, nodes {\em adopting} or becoming infected with information); they {\em ignore the review text}.

\noindent {\em (f) NetInfluence}: This model~\cite{NetInfluence} learns the probability of one node influencing another based on logs of their past propagation ({\em action logs}). The model assumes that when a user $u$ in a network is influenced to perform an action, it may be influenced by its neighbors ($\Psi_u$ in our setting) who have performed the action before. Therefore, each of these predecessors share the ``credit" for influencing $u$ to perform that action. In order to adapt their model to our setting, we consider the event of writing a review on an item $i$ to be an {\em action} at a given timestamp. 
 Therefore input is the set of actions $\langle u, i, t \rangle$ and $\langle u, v \rangle$. Although the authors do not perform recommendation, we use their estimated ``influence'' scores to construct $\Psi$ (refer to Equation~\ref{eq:psi}).\footnote{We used their code available at {http://www.cs.ubc.ca/~goyal/code-release.php/}}. This allows us to use all the features (F3) in Section~\ref{sec:item-rating} derived from the influence network in addition to the rating bias features (F1). Since they do not model text, the language model features are not applicable.

\noindent {\bf (D) GhostLink}: We evaluate GhostLink  with various combinations of the feature sets. In particular, we consider: (a) rating bias (F2, F3.i), (b) network influence (F3), (c) combining rating and network influence (F2, F3), (d) language model (F1), and the full model (F1, F2, F3).


\begin{table}[!t!]
	{\small
		\begin{tabular}{p{4.2cm}cccc}
			\toprule						
			{\bf } &	{\bf Beer}	& {\bf Rate} & {\bf \hspace*{-3mm}Amazon\hspace*{-3mm}} & {\bf \hspace*{-3mm}Amazon\hspace*{-3mm}}\\
			{\bf Models} &	{\bf \hspace*{-3mm}advocate\hspace*{-3mm}}	& {\bf beer} & {\bf Foods} & {\bf Movies}\\\midrule
			{\bf (D) GhostLink}	& {\bf 0.282}	& {\bf 0.250}	& {\bf 0.711}	& {\bf 0.646} \\
			{(Rating + Network + Time + Text)}	& 	& 	& &  \\
			Rating Bias & 0.458	& 0.376	& 1.245	& 1.062 \\
			Network Influence  & 0.443	& 0.386	 & 1.652 & 1.434\\
			Rating + Network Influence &	0.433 &	0.347 &	1.236 &	1.050\\
			Language Model & 1.069	& 1.148	& 3.481 & 4.427 \\\midrule
			{\bf (B) Rating + Text-aware} & & & &\\
			Text-based Collab. Filtering~\cite{mcauleyrecsys2013, rosenzviUAI2004} &	0.373 & 0.302 & 1.347 &	1.233\\\midrule
			 {\bf (C) Rating + Time + Network{\scriptsize-aware}} & & & &\\
			 {NetInfluence}~\cite{NetInfluence} & 0.465 & 0.426 & 0.93 & 0.878\\\midrule
			 {\bf (A) Rating + Time-aware} & & & &\\
			{LFM}~\cite{KorenKDD2010}& 0.559 & 0.917 & 1.465 & 1.620 \\
			{Community at uniform rate}~\cite{McAuley2013,xiong2010temporal,Xiang2010} & 0.582 & 0.945 & 1.530 & 1.727\\
			{User at uniform rate}~\cite{McAuley2013} & 0.586 & 0.950 & 1.523 & 1.729\\
			{Community at learned rate}~\cite{McAuley2013} & 0.532 & 0.833 & 1.529 & 1.729\\
			{User at learned rate}~\cite{McAuley2013}& 0.610 & 0.797 & 1.007 & 0.891\\
			\bottomrule
		\end{tabular}
		\caption{Mean squared error for rating prediction (lower is better). GhostLink outperforms competing methods. }
		\label{tab:MSE}
		\vspace{-3em}
	}
\end{table}

\noindent {\bf Results:} Table~\ref{tab:MSE} shows the results. Standard latent factor collaborative filtering models and most of its temporal variations (Models: A) that leverage rating and temporal dynamics but ignore text and network influence perform the worse.
We observe that the network diffusion based model that incorporates the latent influence network from temporal traces in addition to the rating information (Models: C) perform much better than the previous models not considering the network information. However, these too ignored the textual signals. Finally, we observe that contextual information harnessed from the review content in addition to rating information (Models: B) outperforms all of the previous models.

From the variations of Ghostlink (using only language model), we observe that textual features alone are not helpful. GhostLink progressively improves as we incorporate more influence-specific features. Finally, the joint model leveraging all of context, rating, temporal and influence features incurs the least error. Comparison with the best performing baseline models shows the power of combined contextual and influence features over only context (Models: B) or only network influence (Models: C). 

\noindent {\bf Additional Network-aware models}: We also explored NetInf~\cite{NetInf} for tracing paths of diffusion and influence through networks. Given the times when nodes adopt pieces of information or become infected, NetInf identifies the optimal network that best explains the observed infection times.  
To adapt NetInf to our setting, we consider all the reviews on an item to form a cascade --- with the total number of cascades equal to the number of items. For each cascade (item), the input is the set of reviews on the item by users $u$ at timestamps $t$. However,  NetInf yielded extremely sparse networks on our datasets --- suffering from multiple modeling assumptions like a single influence point for a node in a cascade, static propagation and fixed transmission rates for all the nodes. For example, in BeerAdvocate it extracted only $5$ pairs of influenced interactions. In contrast, both our model as well as NetInfluence, the influence probability $\Psi_{u,v}$ varies for every pair of nodes.




\subsection{Facet Preference Divergence}
In this study, we want to examine if there is any difference between the latent facet preference of users, as opposed to their \emph{observed} preference, and their preference when acting as an influencer (see Sec.~\ref{sec:network} for the definition of these preference distributions).

 {\small
	\begin{table}
		\centering
		\setlength{\tabcolsep}{1pt}
		\begin{tabular}{p{1.9cm}|p{2.1cm}|p{2.2cm}|p{1.9cm}}
			\toprule						
			{\bf Dataset } &	{ C1:\ {\scriptsize$\theta^{obs}_u$ vs. $\theta^{latent}_u$}}	& { C2:\ {\scriptsize $\theta^{latent}_u$ vs.  $\theta^{infl}_u$}} & { C3:\ {\scriptsize $\theta^{obs}_u$ vs. $\theta^{infl}_u$}}\\\midrule
			Beeradvocate & 0.318	& 0.067	& 0.315\\
			Ratebeer & 0.483	& 0.067	& 0.328\\
			Amazon Foods & 0.368	& 0.202	& 0.429\\
			Amazon Movies & 0.370& 0.110 & 0.321\\
			\bottomrule
		\end{tabular}
		\caption{Facet preference divergence between distributions.}
		\label{tab:JSD}
		\vspace{-2em}
	\end{table}
}

We compute the Jensen-Shannon Divergence (JSD) between the different distributions $\theta^{infl}_u, \theta^{latent}_u, \theta^{obs}_u$ to observe their difference. JSD is a symmetrized form of  Kullback-Leibler divergence that is normalized between $0$ - $1$ with $0$ indicating identical distributions. 
%
We compute the JSD results averaged over all the users $u$ in the community, i.e.
$\frac{1}{|U|} \sum_u JSD(\theta^{x}_u \ || \ \theta^{y}_u)$  with the corresponding $x, y \in \{latent, infl, obs\}$.  Table~\ref{tab:JSD} shows the  results.

\noindent We observe (statistically) significant difference between the latent facet preferences of users from that observed/acquired in a community (C1).
This result indicates the strong occurrence of social influence on user preferences in online communities.
We also find that users are more likely to use their original latent preferences to influence others in the community, 
rather than their acquired ones. 
That is, the JSD between the influencer and latent  facet preference distribution (C2) is always significantly smaller than the JSD between the observed and the influencer distribution (C3).

\subsection{Finding Influential Members}

GhostLink generates a directed, weighted influence network $G = (U, E)$ using the user-influencer distribution $\Psi$. Given such a network, we can find influential nodes in the network. We used several algorithms to measure authority like Pagerank, HITS, degree centrality etc. out of which eigenvector centrality performed the best\footnote{Note that these baselines already subsume simpler activity based ranking (e.g., based on number of reviews written).}. The basic idea behind eigenvector centrality is that a node is considered influential, not just if it connects to many nodes (as in simple degree centrality) but if it connects to high-scoring nodes in the network. 
%
%
Given the eigenvector centrality score $x_v$ for each node $v$, we can compute a ranked list of users.

{\bf Comparison: } 
An obvious question is whether this ranking based on the influence graph $\Psi$ is really helpful? Or put differently: Does this perform better compared to a simpler graph-based influence  measure? A natural choice, for example, would be the temporal co-reviewing behavior of users. 
To construct such a graph, we can connect two users $u$ and $v$ with a directed edge if $u$ writes a review following $v$. The weight of this edge corresponds to all such reviews (following the above temporal order) aggregated across all the items. Therefore, $v$ acts as an influencer if $u$ closely follows his reviews. We choose a cut-off threshold of at least $5$ reviews. Also for this graph, we can compute eigenvector centrality scores, and obtain a ranked list of users as described above. 

{\bf The task: }  We want to find which of the above graphs gives a better ranking of users. We perform this experiment in the Beeradvocate and Ratebeer communities. In these communities, users are awarded points based on factors like: their community engagement, how other users find their reviews helpful and rate them, as well as their expertise on beers. This is moderated by the community administrators. For instance, in Beeradvocate users are awarded Karma points\footnote{https://www.beeradvocate.com/community/threads/beer-karma-explained.184895/}. The exact algorithm for calculation of these points is not made public to users as it can be game to manipulation -- and of course, these scores are also not used in GhostLink.

{\small
	\begin{table}
		\centering
		\begin{tabular}{lcc}
			\toprule						
			{\bf Dataset} & {\bf Model} & {\bf Pearson Correlation}\\\midrule
			Beeradvocate  & {\bf GhostLink} & {\bf 0.708}\\
			& NetInfluence & 0.616\\
			& Temporal co-reviewing & 0.400\\\midrule
			Ratebeer & {\bf GhostLink}  & {\bf 0.736}\\
			& NetInfluence & 0.653\\
			& Temporal co-reviewing & 0.615\\
			\bottomrule
		\end{tabular}
		\caption{Pearson correlation (higher is better) between different models to find influential users in the community.}
		\label{tab:cor}
		\vspace{-2em}
	\end{table}
}

We used these points as a proxy for user authority, and rank the users. This ranked list is used as a reference list (ground-truth) for comparison. That is, we use the ranked list of users based on eigenvector centrality scores from our influence graph, and \text{compute} Pearson correlation with the reference list\footnote{Other ranking measures (Kendall-Tau, Spearman Rho) yield similar improvements.}. A correlation score of $1$ indicates complete agreement, whereas $-1$ indicates complete disagreement. We can also do the same for the ranked list of users based on their co-reviewing behavior. As another strong baseline, we also consider the influence scores for the users as generated by NetInfluence~\cite{NetInfluence}.
Table~\ref{tab:cor} shows the results.

{\small
	\begin{table}[!t!]
		\centering	\setlength{\tabcolsep}{1.4mm}
		\begin{tabular}{|c|cc|cc|cc|}
			\toprule						
			{\bf Dataset} &   \multicolumn{2}{c|}{{\bf IG}} & \multicolumn{4}{c|}{{\bf MWSF}}\\
			& {Edges} & {Weight}  & {Edges} & {\% of IG} & {Weight} & {\% of IG}\\\midrule
			Beeradvocate &   180.5K & 31.8K & 132.5K & 73.40\%  & 31.6K & 99.37\% \\
			Ratebeer	&  152.8K& 24.7K  & 95.4K & 62.43\% & 24.5K & 99.19 \%\\					
			Amazon Foods &  107.5K & 59.51K & 104.1K & 96.84 \% & 59.47K & 99.93\% \\
			Amazon Movies &  589K &  145.4K & 476K & 80.81\% & 145K & 99.72\% \\																	
			\bottomrule
		\end{tabular}
		\caption{Structure of the latent influence networks: The Influence Graph (IG) is well represented by a Maximum Weighted Spanning Forest (MWSF) 
		}
		\label{tab:structure}
		\vspace{-1.5em}
	\end{table}
}

\begin{figure*}[!t!]
	\centering
	\includegraphics[width=\linewidth, height=4.5cm]{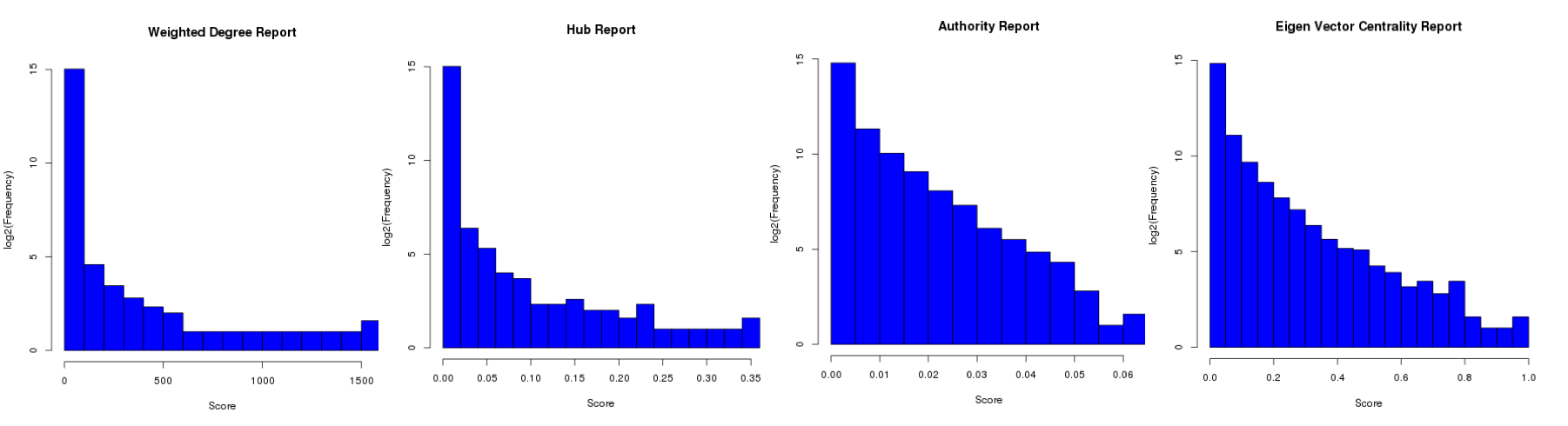}
	\vspace{-2.5em}
	\caption{Distribution of nodes with scores (weighted degree, hubs, authorities, and eigen vector centralities in order from left to right) in log-scale for the extracted influence graph (note the change in scale of scores for each figure) in Beeradvocate data.}
	\vspace{-1em}
	\label{fig:graph-stat}
\end{figure*}

\begin{figure}[!b!]
	\vspace{-1em}
	\centering
	\includegraphics[width=\linewidth, height=4.5cm]{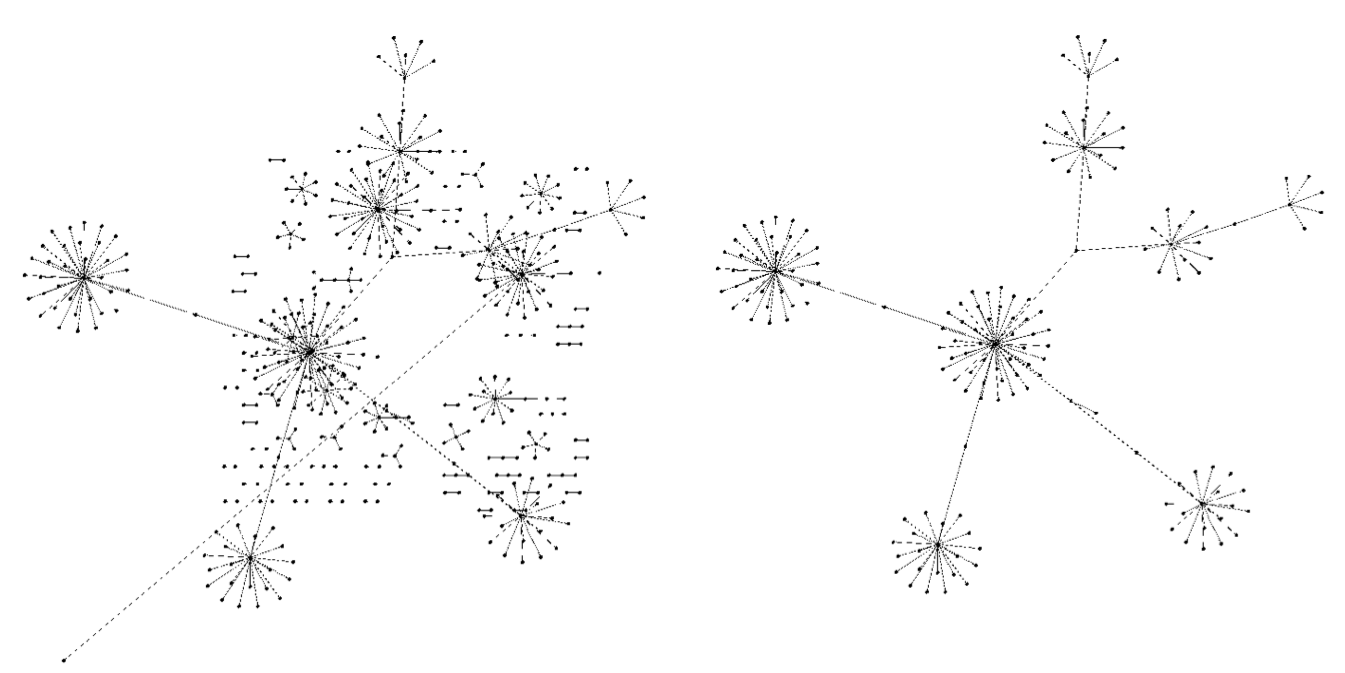}
	\vspace{-1em}
	\caption{\small Maximum Weighted Spanning Forest corresponding to a representative facet (left); and its giant component (right).}
	\label{fig:facet-infl}
\end{figure}

We observe that the ranking computed with our influence graph performs much better (higher correlation with ground-truth) than the temporal co-reviewing baseline. Thus, the learned influence network indeed captures more information than simple co-review\-ing behavior and even the more advanced diffusion based NetInfluence model, and enables us to find influential users better. Note again, that the point-based scores used for the ground-truth ranking have {\em not been used} in GhostLink. 

\subsection{Structure of the Influence Network}


Last, we analyze the structure of the influence network $\Psi$. Our first research question is: How is the mass (sum of influence/edge weights) distributed in the network? Is it randomly spread out, or do we observe any particular structure (e.g., resembling a tree-like structure).
For this, we computed a Maximum Weighted Spanning Tree from the graph 
(or spanning forest as the graph is not connected) and computed the sum of its edge-weights, i.e. its mass.

 Table~\ref{tab:structure} shows the statistics of the constructed MWSF over different datasets and compares it with the (original) influence graph (IG). We observe that {\em the majority of mass of the influence graph is concentrated in giant tree-components}. For example, in Beeradvocate, $99.37\%$ of the mass of the influence graph is concentrated in the MWSF. The forest accounts for $73.40\%$ of the edges in the influence graph. Thus, the remaining $26.6\%$ of the edges contribute only marginally, and can be pruned out. 
 This tree-like influence matches intuition: a user often influences many other users, while she herself gets primarily influenced by a few -- surprisingly, in the majority of cases only by a single other user -- as indicated by the good approximation of the graph via a tree (preservation of mass). Figure~\ref{fig:facet-infl} shows the MWSF for a representative facet ``yuengling", and its giant component.

 The tree structure in Figure~\ref{fig:facet-infl} shows another characteristics: only a few users seem to influence many others (it resembles a snowflake) in the community. This brings us to our second research question: Do we observe --- similar to real-world networks --- specific power-law behaviors? For example, are the majority of nodes `influencees', and only a few nodes are `influencers'?
 
Figure~\ref{fig:graph-stat} analyzes this aspect. Here we illustrate the distribution of nodes with weighted degree, hub \& authority, and eigen vector centrality scores for our influence graph plotted in {\em log-scale}. These statistics are for the Beeradvocate community. The statistics for other communities are similar.
 Indeed, we observe power-law like distributions with many influencees and a few influencers. 

For the HITS algorithm, a hub --- with a lot of outgoing edges --- is a user who influences a lot of other users; whereas an authority --- with a lot of incoming edges --- is the one getting influenced by other influential users. Note that each node can be a hub and an authority with different scores simultaneously. We observe that there are a lot of hubs (influencers) with very low influence scores, and only few with very high influence. From the authority report, we see that there are less number of incoming edges to nodes (note the really small range of authority scores of nodes). This indicates that users generally get influenced by only a few users in the community --- confirming the tree-like structure of the influence graph.

\section{Related Work}

\todo{this sentence is broken!!! and some references are missing.. probably you did not push the bib?}State-of-the-art recommender systems  exploit user-user and item-item similarities using latent factor models \cite{korenKDD2008, koren2011advances}. Temporal patterns in ratings such as bursts, bias, and anomalies 
are studied in \cite{KorenKDD2010, XiangKDD2010, Gunnemann2014}. Recent works~\cite{mcauleyrecsys2013, wang2011, mukherjeeSDM2014} have further considered review texts for content-aware recommender systems. However, all of these works assume that users participate independently in the community which is rarely the case.

Social-aware recommender systems~\cite{Tang:2012:MDM:2124295.2124309, Tang:2013:ELG:2540128.2540519, DBLP:journals/datamine/LiuTHY12,DBLP:conf/sdm/ZhangYWSZ17, DBLP:journals/ida/MeiYSM17, DBLP:journals/jidm/FelicioPAAP16, Ye:2012:ESI:2348283.2348373, 
	7944514, Krishnan:2010, 
	 DBLP:conf/ecai/HuangCGSY10} 
exploit peers and friends of users to extract more insights from their activities, likes, and content sharing patterns using homophily. 
In absence of explicit social networks in many communities, some  works~\cite{Guo:2014:RTE:2554850.2554878, Lin:2014:PNR:2535053.2535249, Ma:2013:ESI:2484028.2484059, Ma:2011:RSS:1935826.1935877} exploit collaborative filtering to extract implicit social relationships based on the historical rating behavior. Some of these works also leverage signals like pre-defined trust metrics, and partial or explicit social links. \cite{Lin:2014:PNR:2535053.2535249, DBLP:conf/icdm/ZhangWZLTZ16} use time as an additional dimension along with ratings. 

Information diffusion based works~\cite{NetInf,Connie,NetRate,NetInfluence} that model underlying {\em latent} influence or diffusion networks do not consider text. Some of them have strong assumptions in terms of known transmission rates, static and homogeneous transmission etc. 
Recent works on text-based diffusion~\cite{Wang2014, Du2013, HawkesTopic} alleviate some of these assumptions. However, they also make some assumptions regarding topics of diffusion being known, network being explicit etc. Most importantly, none of these works are geared for item recommendation and do not study the characteristics of review communities.


Works in modeling influence in heterogeneous networks~\cite{DBLP:journals/datamine/LiuTHY12} and citations networks~\cite{DBLP:conf/icml/DietzBS07} assume the presence of explicit user-user links. Prior works on modeling influence propagation and cascades~\cite{Myers:2012:IDE:2339530.2339540} also consider a given network to propagate influence scores. 
Learning a latent influence network has been possible in the field of information propagation when observing cascades of events \cite{DBLP:journals/tkdd/Gomez-RodriguezLK12,DBLP:conf/icdm/ZhangWZLTZ16}. However, these works have not considered the setting where only review text is \todo[color=green]{Subho, it would be good if you look in the paper \cite{DBLP:conf/icdm/ZhangWZLTZ16}. I guess they have a good RW. We might want to cite some of their further papers. Just that no reviewer complains!} available, and no explicit networks.

{\em In contrast to prior works, GhostLink learns the latent influence network solely from timestamped user reviews, without requiring any explicit user-user link/rating information}.  It uses this network to improve item rating prediction considering implicit social influence.

\vspace{-0.5em}
\section{Conclusion}

We presented GhostLink, an unsupervised generative model to extract the underlying influence graph in online communities dealing with items of fine taste like movies, food and beer without requiring any explicit user-user links or ratings. Given only timestamped reviews of users, we leverage opinion conformity from overlapping facet descriptions in co-reviewed content and their temporal traces to extract this graph. Furthermore, we use this influence network to improve item rating prediction by $23\%$ over state-of-the-art methods by capturing implicit social influence. We show in large-scale experiments in four real-life communities with $13$ million reviews that GhostLink outperforms several state-of-the-art baselines for tasks like recommendation and identifying influential users. 

\textbf{Acknowledgements.}
This research was supported by the German Research Foundation, Emmy Noether grant GU 1409/2-1. 

We would like to sincerely thank Christos Faloutsos for his insightful and constructive comments on the paper.

\bibliographystyle{ACM-Reference-Format}
\bibliography{infl} 

\end{document}